\documentclass{article}

\usepackage{epsf}
\usepackage{epsfig}

\def\nn{\nonumber}

\def\la{\langle}
\def\ra{\rangle}

\def\l{\left}
\def\r{\right}
\def\nn{\nonumber}

\def\beq{\begin{equation}}
\def\eeq{\end{equation}}
\def\bea{\begin{eqnarray}}
\def\eea{\end{eqnarray}}
\def\barr{\begin{array}}
\def\earr{\end{array}}

\begin{document}

\begin{titlepage}
\begin{flushright}
ROMA-1393/04\\
SHEP 0436\\
\end{flushright}
\vskip 0.5cm
\begin{center}
{\Large \bf Twisted Boundary Conditions in Lattice Simulations}
\vskip1cm {\large\bf
C.T.~Sachrajda$^{a}$ and G.~Villadoro$^b$ } \\
\vspace{.5cm} {\normalsize {\sl $^a$ School of Physics and
Astronomy, Univ. of Southampton,\\ Southampton, SO17 1BJ, UK. \\
\vspace{.2cm} $^b$ Dip. di Fisica, Univ. di Roma ``La Sapienza"
and INFN,\\ Sezione di Roma, P.le A. Moro 2, I-00185 Rome,
Italy.}}

\vskip1.0cm {\large\bf Abstract\\[10pt]} \parbox[t]{\textwidth}{{
By imposing twisted boundary conditions on quark fields it is
possible to access components of momenta other than integer
multiples of $2\pi/L$ on a lattice with spatial volume $L^3$. We
use Chiral Perturbation Theory to study finite-volume effects with
twisted boundary conditions for quantities without final-state
interactions, such as meson masses, decay constants and
semileptonic form factors, and confirm that they remain
exponentially small with the volume. We show that this is also the
case for \textit{partially twisted} boundary conditions, in which
(some of) the valence quarks satisfy twisted boundary conditions
but the sea quarks satisfy periodic boundary conditions. This
observation implies that it is not necessary to generate new gluon
configurations for every choice of the twist angle, making the
method much more practicable. For $K\to\pi\pi$ decays we show that
the breaking of isospin symmetry by the twisted boundary
conditions implies that the amplitudes cannot be determined in
general (on this point we disagree with a recent claim). }}
\end{center}
\end{titlepage}

\section{Introduction}
In lattice simulations of QCD on a cubic volume ($V=L^3$) with
periodic boundary conditions imposed on the fields, the hadronic
momenta ($p$) are quantized according to $p_i=2\pi/L\times n_i$,
where $i=1,2,3$ and the $n_i$ are integers. For currently
available lattices this means that the lowest non-zero momentum is
large (typically about 500\,MeV or so) and there are big gaps
between neighbouring momenta. This limits the phenomenological
reach of the simulations. In ref.\,\cite{bedaque} Bedaque proposed
the use of non-periodic boundary conditions which would allow
hadrons with arbitrarily small momenta to be simulated (see also
the references cited in \cite{bedaque} for earlier related ideas).
We refer to these boundary conditions as \textit{twisted} boundary
conditions~\footnote{An analogous method was already introduced
many years ago in the context of extra-dimensions \cite{ss} and is
still widely used for breaking spontaneously some of the action
symmetries. The breaking is spontaneous since it is caused by a
non-local effect.}. This technique has subsequently been used in a
quenched study of the energy-momentum dispersion relations of
pseudoscalar mesons~\cite{dpt} and the finite-volume corrections
for two-particle states with twisted boundary conditions have been
calculated~\cite{dt}.

In this letter we use chiral perturbation theory ($\chi$PT) to
analyse some of the properties of twisted boundary conditions and
show that:
\begin{enumerate}
\item For physical quantities without final state interactions,
such as masses or matrix elements of local operators between
states consisting of the vacuum or a single hadron, the flavour
symmetry breaking induced by the twist only affects the
finite-volume corrections, which nevertheless remain exponentially
small.\label{item:fv}
\item For amplitudes which involve final-state interactions, such
as those for $K\to\pi\pi$ decays, in general it is not possible to
extract the physical matrix elements using twisted boundary
conditions (see sec.\,\ref{sec:fveffects}). On this point we
disagree with ref.\,\cite{dt}\label{item:fsi}.
\item For \textit{Partially Twisted} boundary conditions, in which
(some of) the valence quarks satisfy twisted boundary conditions
but the sea quarks satisfy periodic ones, one also obtains the
physical quantities described in item\,\ref{item:fv} with
exponential precision in the volume. This implies that in
unquenched simulations it is not necessary to generate new gluon
configurations for every choice of boundary condition, thus making
the method much more practicable.
\end{enumerate}
In ref.\,~\cite{kc} Kim and Christ propose H- and G-parity
boundary conditions in which the minimum non-zero hadronic momenta
are reduced from $2\pi/L\to\pi/L$ (see also ref.\,\cite{wiese}).
These authors impose H-parity boundary conditions for $K\to\pi\pi$
decays in which the two-pions are in an $I=2$ state. This is a
particular case of twisted boundary conditions, corresponding to
the specific choice of $\pi$ for the twisting angle (as stated in
item\,\ref{item:fsi} above and explained in
sec.\,\ref{sec:fveffects} below, it is not possible to study
$K\to\pi\pi$ decays with a general choice of twisting angle). Kim
and Christ also show that G-parity boundary conditions, which
exploit the discrete charge conjugation transformations, can be
used for an $I=0$ two-pion state (in unquenched simulations), but
the formalism will have to be extended to incorporate the kaon.
Although we do comment below on H- and G-parity boundary
conditions in order to illustrate our discussion, the main focus
of this letter is on boundary conditions based on continuous
symmetries.

When considering $K\to\pi\pi$ decays, throughout this letter we
restrict our discussion to the centre-of-mass frame for the two
pions. For this case and with periodic boundary conditions, the
finite volume corrections which decrease as powers of the volume
have been derived for the two-pion spectrum~\cite{ml} and matrix
elements~\cite{ll,lmst}. At present the theory of finite-volume
corrections in a moving frame has not been developed for matrix
elements (but for a discussion of finite volume corrections to the
two-pion spectrum in a moving frame see ref.\,\cite{rg}). We
therefore do not generalise our discussion to the moving frame at
this stage.

The plan of the remainder of this letter is as follows. In the
following section we define twisted boundary conditions in QCD and
briefly review their properties. In sec.\,\ref{sec:chiL} we impose
twisted boundary conditions on the chiral lagrangian and
demonstrate that their effect is to shift the momenta of internal
propagators and external mesons by amounts corresponding to the
twists. Sec.~\ref{sec:fveffects} contains a discussion of
finite-volume effects when twisted boundary conditions have been
imposed. We discuss partially twisted boundary conditions in
sec.~\ref{sec:partiallytwisted} and present our conclusions in
sec.\,\ref{sec:concs}. There are two appendices in which we derive
the finite-volume corrections with twisted boundary conditions at
one-loop order in $\chi$PT (appendix\,\ref{app:tadpoles}) and
present the corresponding results for meson masses and decay
constants (appendix\,\ref{app:fvfull}).

\section{Twisted Boundary Conditions in QCD}\label{sec:gbc}
In this section we will define the twisted boundary conditions and
derive some of the constraints they have to satisfy. Since the
choice of boundary conditions is a non-local effect, we can
present the discussion, without any loss of generality, within the
framework of continuum quantum field theory. It should be noted
however, that the discussion also applies to every lattice
discretization. Although local discretization artefacts may affect
the constant pre-factors, they do not affect the functional
behaviour with the volume. For definiteness we present the
discussion in Euclidean space with an infinite time dimension and
a finite cubic spatial volume of size $L^3$.

When formulating quantum field theory in a finite cubic volume, in
order to avoid boundary terms, periodic boundary conditions are
frequently imposed on the fields. This is equivalent to defining
the theory on a torus and the periodicity of the fields ensures
that the fields are single valued. However requiring that the
fields be single valued is not necessary; it is sufficient instead
to require that the observables be single valued, which is
equivalent to the condition that the action be single valued on
the torus. This means that the generic field $\phi$ has to respect
the following boundary conditions:
\begin{equation}
\phi(x_i+L)=U_i\,\phi(x_i)\qquad i=1,2,3\,, \label{eq:bc}
\end{equation} where the index $i$ is not summed and $U_i$ is a
symmetry of the action. Imposing the condition in
eq.\,(\ref{eq:bc}) is sufficient to cancel the boundary terms.

Consider now the Dirac term in the (Euclidean) QCD Lagrangian,
\beq {\cal L}=\bar q(x) \l (\,\slash \hspace{-7pt}D + M \r )
q(x)\,,\label{eq:freeL} \eeq where for our discussion it will be
convenient to consider the quark field $q(x)$ to be a vector in
flavour space and the quark mass matrix $M$ to be a diagonal
matrix. The possible boundary conditions depend on the symmetries
of the action, and in particular on the form of $M$, i.e. on
whether there is any degeneracy assumed for the quark masses. Here
we will consider the most general continuous symmetry, i.e. the
flavour symmetry group $U(N)_V$ and its subgroups, and will not
discuss the possible use of discrete symmetries (and charge
conjugation in particular~\cite{kc}). An advantage of the use of
continuous symmetries is that the minimum momentum can take any
value less that $2\pi/L$, whereas with discrete symmetries such as
G-parity the lowest momentum is $\pi/L$. Eq.\,(\ref{eq:bc}) then
implies that $U_i$ has to commute with the Dirac operator, and in
particular with the quark mass matrix. For general values of the
quark masses this implies that $U_i$ is a diagonal matrix. In the
isospin limit one could in principle, take $U_i$ to be
non-diagonal in the $u$\,--\,$d$ sector, however this choice
breaks the conservation of electric charge, and will not be
considered explicitly here. We can therefore write the boundary
condition for the quark fields in the form:\beq
q(x_i+L)=U_i\,q(x_i)=\exp{(i\,\theta_i^a
T^a)}\,q(x_i)\equiv\exp{(i\,\Theta_i)}\,q(x_i)\,,
\label{eq:defUmu} \eeq where the $T^a$'s are the generators in the
Cartan subalgebra of the flavour $U(N)_V$ group commuting with the
quark mass matrix. It is convenient to redefine the quark fields
by:
\begin{equation}
q(x)\equiv V(x)\,\tilde q(x) \qquad\textrm{where}\qquad
V(x)\equiv\exp \l (i\,\frac{\Theta_i}{L}
x_i\r)\,.\label{eq:defUx}\end{equation} The fields $\tilde{q}(x)$
satisfy periodic boundary conditions,
\begin{equation} \tilde q(x_i+L)=\tilde q(x_i)
\,, \end{equation} and the Lagrangian (\ref{eq:freeL}) is given in
terms of these fields by:
\begin{equation} {\cal L}=\bar{\tilde{q}}(x) \l (\,\slash \hspace{-7pt}D +
    (V^\dagger(x) \slash \hspace{-5pt} \partial V(x)) + M \r ) \tilde q(x)
    =\bar {\tilde q}(x) \l (\,\slash \hspace{-7pt} \tilde{D} + M \r ) \tilde q(x)\,,
\end{equation}
where
\begin{equation}
\tilde{D}_\mu=D_\mu+iB_\mu \qquad\textrm{where}\qquad
B_i=\frac{\Theta_i}{L}\,\quad\textrm{for }i=1,2,3\textrm{\quad
and}\quad B_4=0\,.\end{equation} This is the Lagrangian of QCD
with quark fields satisfying periodic boundary conditions
interacting with a constant background vector field which couples
to quarks with charges determined by the phases in the twisted
boundary conditions. The external field, in addition to breaking
the cubic group symmetry, breaks also all the symmetries which do
not commute with it. For generic diagonal $B_i$ the broken
symmetries are flavour SU(3) and $I^2$, but not $I_z$, strangeness
and the electric charge.

To illustrate some of the above points and the kinematic nature of
the symmetry breaking induced by the twisted boundary conditions,
we end this section by exhibiting the propagator of a free quark
using both the $q$ and $\tilde{q}$ definitions. For compactness of
notation we drop the flavour index and take $B=\theta/L$ to be the
twist corresponding to the flavour represented by $q$ with mass
$M$. The propagators are then
\begin{eqnarray}
S(x)&\equiv&\langle\,q(x)\,\bar{q}(0)\,\rangle=\int\,\frac
{dk_4}{2\pi}\frac{1}{L^3}\sum_{\vec{k}}\,\frac{e^{i(k+B)\cdot x}}
{i(\slash\hspace{-5pt}k+\slash\hspace{-7pt}B)+M}=e^{iB\cdot x}\tilde{S}(x)\\
\textrm{and}\qquad\tilde{S}(x)&\equiv&
\langle\,\tilde{q}(x)\,\bar{\tilde{q}}(0)\,\rangle=\int\,\frac
{dk_4}{2\pi}\frac{1}{L^3}\sum_{\vec{k}}\,\frac{e^{ik\cdot x}}
{i(\slash\hspace{-5pt}k+\slash\hspace{-7pt}B)+M}\label{eq:sstilde}
\end{eqnarray}
where in both cases the sum is over momenta
$\vec{k}=(2\pi/L)\,\vec{n}$ and $\vec{n}$ is a vector of integers.
$S(x)$ satisfies the twisted boundary condition
($S(x_i+L)=\exp(i\,\theta_i)S(x)$) and
$(\slash\hspace{-5pt}\partial+M)S(x)=\delta(x_4)\,\delta^{(3)}_{\vec{x},0}$
whereas $\tilde{S}(x)$ satisfies periodic boundary conditions
($\tilde{S}(x_i+L)=\tilde{S}(x)$) and
$(\slash\hspace{-5pt}\partial+i\,\slash\hspace{-7pt}B+M)\tilde{S}(x)
=\delta(x_4)\,\delta^{(3)}_{\vec{x},0}$\,. The momentum in the
denominators is shifted (or boosted) by $\theta/L$.

\section{The Effective Chiral Lagrangian}\label{sec:chiL}
In this section we derive the low-energy effective Lagrangian for
QCD in the presence of twisted boundary conditions and study its
properties. The derivation of the chiral Lagrangian could be
performed directly by coupling the Gasser-Leutwyler Lagrangian to
the external vector field $B_\mu$ introduced above. Here instead,
we choose to follow the steps of sec.~\ref{sec:gbc} in order to
show the equivalence with QCD explicitly. Imposing the boundary
conditions of eq.~(\ref{eq:bc}) on the fields, implies that:
\begin{equation}
\Sigma(x_i+L)= U_i \Sigma(x_i) U_i^\dagger\,,
\end{equation} where $\Sigma$ is the coset representative of
$SU(3)_L \times SU(3)_R /SU(3)_V$ and $U_i$ is defined in
eq.~(\ref{eq:defUmu}). Note that this relation is completely fixed
once the quark boundary conditions are imposed, and the results
below are implied unambiguously by this relation. Following the
presentation in sec.~\ref{sec:gbc}, we redefine the fields by:
\begin{equation} \Sigma(x)\equiv V(x) \tilde \Sigma(x)
V^\dagger(x)\,,\label{eq:Sigmatildedef}\end{equation} so that
$\tilde\Sigma$ satisfies periodic boundary conditions.
Eq.\,(\ref{eq:Sigmatildedef}) corresponds to a local symmetry
transformation so that only the derivative terms are affected:
\bea \partial_\mu \Sigma&=&V(x) (\partial_\mu \tilde \Sigma )
V^\dagger(x)+
    V(x)\l ( V^\dagger(x) \partial_\mu V(x)\r)\tilde
    \Sigma V^\dagger(x)+V(x) \tilde \Sigma \l(\l
    (\partial_\mu V^\dagger(x)\r) V(x)\r) V^\dagger(x) \\
    &=&V(x) \l ( \partial_\mu \tilde \Sigma + [V^\dagger(x) \partial_\mu V(x), \tilde \Sigma] \r)
     V^\dagger(x)\\
     &=&V(x) \l ( \partial_\mu \tilde \Sigma + [iB_\mu, \tilde \Sigma] \r)
     V^\dagger(x)\,.
\eea In terms of $\tilde\Sigma$ the chiral Lagrangian reads \bea
{\cal L}_{\chi PT}&=&\frac{f^2}{8}\left\{\la \tilde{D}^\mu \tilde
\Sigma^\dagger \tilde{D}_\mu \tilde \Sigma \ra
    - \la \tilde \Sigma \chi^\dagger +\chi \tilde \Sigma^\dagger
    \ra\right\}
    \label{eq:chilag} \\
\textrm{where}\quad\tilde{D}_\mu \tilde \Sigma&\equiv&
\partial_\mu \tilde \Sigma + i [B_\mu, \tilde \Sigma]\,,
\label{eq:Dsigma} \eea and $\la\ \ra$ represents the trace. The
Lagrangian in eq.\,(\ref{eq:chilag}) is the standard $\chi PT$
Lagrangian with periodic fields coupled to the vector external
field $B_\mu$. Note that the long-distant nature of the boundary
conditions allows $\chi PT$ to take their effects completely into
account through the simple modification in eqs.\,(\ref{eq:chilag})
and (\ref{eq:Dsigma}). The low energy constants are not affected
by the twist (analogously to the arguments in \cite{gl}).

The effects of the twist on the mesons can be obtained directly
from eq.~(\ref{eq:Dsigma}). The neutral-meson fields commute with
$B_\mu$ (recall that $\Theta$ is diagonal) and do not receive any
boost. The charged-meson fields, on the other hand, are boosted by
an amount proportional to the difference of the twists of the two
valence quarks ($v_1$ and $v_2$):
\beq
[B_i,\sigma^{\pm}]=\l[\frac{\theta_{v_1,i}-\theta_{v_2,i}}{2\,L}\,
\sigma_3\,,\,\sigma^{\pm}\r]= \pm
\frac{\theta_{v_1,i}-\theta_{v_2,i}}{L}\,\sigma^{\pm} \eeq and the
spectrum of allowed meson momenta is shifted accordingly, both in
external states and in internal propagators.

\begin{figure}[h!]
\vspace{0.0in} \centerline{ \epsfxsize=6cm \epsfysize=5.2cm
\epsffile{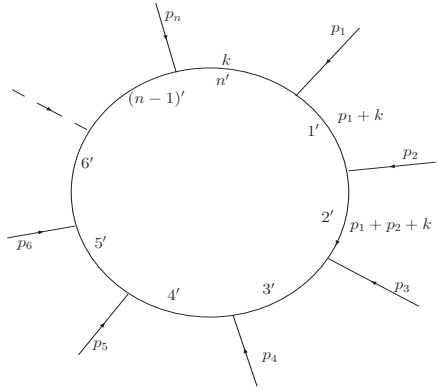}} \caption{Auxiliary one-loop diagram with
$n$ external mesons, used in the demonstration that the effect of
twisting is to shift the internal and external momenta
accordingly. The unprimed and primed variables represent the
external and internal lines respectively.\label{fig:loopdiagram}}
\end{figure}

From the chiral Lagrangian in eq.~(\ref{eq:chilag}), and its
extensions to higher order in the momentum expansion, it follows
that the only effect of the twisted boundary conditions is to
shift all the momenta consistently in order to recover the correct
boost corresponding to the flavour of each propagator and external
line. We illustrate this by considering the loop contribution
represented in fig.\,\ref{fig:loopdiagram}, which may be a
one-loop contribution to an $n$-body process or an insertion in a
higher-order diagram. The contribution from this diagram is of the
form:
\begin{equation}
\int\frac{dk_4}{2\pi}\frac{1}{L^3}\sum_{\vec{k}} \frac{(p_1+\cdots
+ p_r+B_{r'}+k)^\mu \dots}
{[(k+B_{n^\prime})^2+m_{n^\prime}^2]\,[(p_1+B_{1'}+k)^2+m_{1'}^2]\dots
[(p_1+\dots+p_{(n-1)}+B_{(n-1)'}+k)^2+m^2_{(n-1)^\prime}]}\label{eq:oneloopk}\eeq
where the sum is over momenta $k_i=(2\pi/L)\,n_i$ and the
$\{n_i\}$ are integers~\footnote{Note that in general there could
also be an even number of mesons attached to some vertices but
this does not change the validity of the demonstration.}. The
factor in the numerator represents the derivative terms at
vertices in the chiral Lagrangian and $B_{l^\prime}$ refers to the
momentum shift due to the external field on the meson in the
$l^\prime$ propagator of the loop. We now perform the trivial
change of variables $k\to k^\prime=k+B_{n^\prime}$ to rewrite the
sum in eq.\,(\ref{eq:oneloopk}) as
\beq\int\frac{dk^{\prime}_4}{2\pi}
\frac{1}{L^3}\sum_{\vec{k}^\prime} \frac{(p_1+B_1+\cdots +
p_r+B_{r}+k^\prime)^\mu
\dots}{[k^{\prime\,2}+m_{n^\prime}^2]\,[(p_1+B_{1}+k^\prime)^2+m_{1'}^2]\dots
[(p_1+B_1+\dots+p_{(n-1)}+B_{(n-1)}+k^\prime)^2+m_{(n-1)'}^2]}\,,
\label{eq:oneloopkprime}\eeq where now the sum is over momenta
$k^\prime_i=(2\pi/L)\,n_i+B_{n^\prime\,i}$ with integer $n_i$.
$B_i$ is the twist corresponding to the flavour of the external
line $i$ and we have used the fact that at each vertex the sum of
the twists is zero (e.g. $B_{1^\prime}-B_{2^\prime}+B_2=0$). This
condition is a consequence of the invariance of the action under
the twist transformations. So far we have considered the theory on
a single volume, where there are finite-volume artefacts, and in
sec.\,\ref{sec:fveffects} below we investigate the size of these
corrections (which do depend on the boundary conditions which have
been imposed). In phenomenological applications we generally wish
to eliminate FV corrections by taking, in principle at least, the
infinite-volume limit, so that the sum in eq.\,(\ref{eq:oneloopk})
goes over into the integral
\begin{equation}\int\frac{d^4k^\prime}{(2\pi)^4}
\frac{(p_1+B_1+\cdots + p_r+B_{r}+k^\prime)^\mu
\dots}{[k^{\prime\,2}+m_{n^\prime}^2]\,[(p_1+B_{1}+k^\prime)^2+m_{1'}^2]\dots
[(p_1+B_1+\dots+p_{(n-1)}+B_{(n-1)}+k^\prime)^2+m_{(n-1)'}^2]}\,.
\label{eq:integralkprime}\end{equation} As required, this is
precisely the expression for the contribution from this diagram
with external momenta $p_i+B_i$. For fixed external momenta
($p_i+B_i$ in the notation of eq.\,(\ref{eq:integralkprime})), the
integral is independent of the boundary conditions which have been
used in the finite-volume calculations.

We conclude this section with some brief comments about the way
that the infinite volume limit might be taken in principle. We
start of course by studying a physical quantity in a
finite-volume. For illustration imagine that the process depends
on a component of momentum $p_i$ which is smaller than $2\pi/L$
for a particular lattice simulation and so we introduce a twist
$\theta_i$ for the corresponding flavour in direction $i$. Now we
can envisage taking the infinite volume limit in a number of ways.
For example we may keep $\theta_i$ fixed so that $B_i\to0$ as the
volume is increased. $p_i$ is kept fixed in physical units, and
since $p_i=(2\pi/L)\,n_i\,+\,\theta_i/L$ for some integer $n_i$,
as we increase the volume we take higher excitation levels $n_i$.
The effect of the twist decreases as the volume increases, and the
results approach those obtained with periodic boundary conditions.
This is also true for momentum sums such as that in
eq.\,(\ref{eq:oneloopk}), which are dominated by momenta of order
of some physical scales and hence the relevant $n_i$ increase as
$L$ increases. Thus again we see that the effects of the twist
decrease as the volume is increased. This feature is generally
true as long as the infinite-volume limit is taken \textit{keeping
the physics fixed}.

\section{Finite Volume effects with Twisted Boundary Conditions}
\label{sec:fveffects}

Finite-volume corrections in general, and those due to the choice
of boundary conditions in particular, are long-distance effects
which can be studied using $\chi$PT (for sufficiently light
pseudo-Goldstone mesons and large volumes). We start by
considering processes without any final-state interactions, such
as particle masses or matrix elements of local operators with
external states which consist of either the vacuum or a single
hadron. For these quantities finite-volume corrections are known
to be exponentially suppressed with the volume, due to the fact
that in the absence of branch cuts (which is the case for these
quantities), the Poisson formula allows us to replace the sums
over the discrete momenta in finite volume by infinite-volume
integrals. Differences between the two are exponentially small in
the volume and this remains true in the presence of twisted
boundary conditions. As shown in appendix\,\ref{app:tadpoles}, the
finite-volume correction can be calculated in terms of
elliptic-$\vartheta$ functions, and decrease exponentially at
large volumes.

We now report the asymptotic finite-volume corrections (in the
limit $L\to\infty$) for pion masses and decay constants with
twisted boundary conditions; the results demonstrate explicitly
the isospin breaking at finite volume. For each physical quantity
we present the results in the form
\begin{equation}
\frac{\Delta X}{X}\equiv\frac{X(L)-X(\infty)}{X(\infty)}\,,
\label{eq:xdef}\end{equation} where $X(L)$ and $X(\infty)$ are the
results in finite and infinite volume respectively. The full
expressions for the finite-volume corrections (at NLO in $\chi$PT)
can be found in appendix\,\ref{app:fvfull}, and their asymptotic
behaviour as $L\to\infty$ is as follows: \bea \frac{\Delta
m_{\pi^{\pm}}^2}{m_{\pi^{\pm}}^2}&\to&3
\frac{m_\pi^2}{f_\pi^2}\frac{e^{-m_\pi L}}
{(2\pi m_\pi L)^{3/2}} \,, \nn \\
\frac{\Delta m_{\pi^{0}}^2}{m_{\pi^{0}}^2}&\to&3 \frac{m_\pi^2}
{f_\pi^2}\frac{e^{-m_\pi L}}
{(2\pi m_\pi L)^{3/2}}
    \l(\frac23 \sum_{i=1}^3 \cos{\theta_i}-1\r) \,, \nn \\
\frac{\Delta f_{\pi^{\pm}}}{f_{\pi^{\pm}}}&\to&-3 \frac{m_\pi^2}
{f_\pi^2}\frac{e^{-m_\pi L}}{(2\pi m_\pi L)^{3/2}}
    \l(\frac13 \sum_{i=1}^3 \cos{\theta_i}+1 \r) \,, \\
\frac{\Delta f_{\pi^{0}}}{f_{\pi^{0}}}&\to&-3 \frac{m_\pi^2}{f_\pi^2}
\frac{e^{-m_\pi L}}{(2\pi m_\pi L)^{3/2}}
    \l(\frac23 \sum_{i=1}^3 \cos{\theta_i}\r) \,.
\eea

For matrix elements involving two or more mesons in the final
state the situation is more complicated: there are both
exponential and power finite-volume corrections. The latter are
parametrically larger (and in most cases numerically larger).
Two-particle energy shifts due to the finite volume,
Lellouch-L\"uscher factors relating finite-volume matrix elements
to physical amplitudes and finite-volume corrections to the
two-particle interpolating operators at the sink which contain
final state interactions, arise as power corrections in the
volume. If twisted boundary conditions affect these terms, then
they inevitably affect unitarity with obvious consequences for the
extraction of the matrix elements.

Consider, for instance, the case of $K\to\pi\pi$ decays, and
imagine that only the $u$-quark is twisted, so that the charged
pions are boosted but not the neutral ones. In such a situation
$I^2$ is no longer a good quantum number, so that the energy
eigenstates are no longer states with a definite isospin I; in
particular they are no longer states with I=0 or I=2 as is the
case when isospin is a symmetry. This can be seen even in the
free-theory. Since we require that the two-pion state is at rest,
at tree-level the lowest energy state is $|\pi^0\pi^0\rangle$ with
both pions at rest and the first excited state is
$|\pi^+\pi^-\rangle$ with the momenta of each of the two pions
$\vec{p}_{\pi^\pm}=\pm\vec{\theta}/L$, where $\vec\theta$ is the
twist on the up-quark. In the interacting theory the presence of
$\pi^+\pi^-\leftrightarrow\pi^0\pi^0$ transitions complicates the
analysis very significantly and, as explained in the next
paragraph, it is not possible to determine physical $K\to\pi\pi$
amplitudes from simulations on finite volumes using twisted
boundary conditions with the power corrections in the volume kept
under control. These issues were not considered in ref.\,\cite{dt}
and we therefore do not agree that the formulae for finite-volume
corrections presented in that paper can be applied to $K\to\pi\pi$
decays.

We now briefly demonstrate the difficulties in studying quantities
involving two-pion states using boundary conditions which break
isospin invariance. Consider, for example, the correlation
functions
\begin{equation}
\langle
0|\pi^0(t)\pi^0(t)\,\sigma(0)|0\rangle\qquad\textrm{and}\quad
\langle 0|\pi^+(t)\pi^-(t)\,\sigma(0)|0\rangle\,,
\end{equation}
where $\sigma$ is some operator which can create two pions from
the vacuum and $\pi^i$ is an interpolating operator for a pion
with charge $i$. $\sigma$ is placed at the origin and we have
taken a Fourier transform at zero momentum of each of the $\pi^i$
fields so that only their time dependence is exhibited (of course
the boundary conditions induce a momentum of $O(\vec{\Theta}/L)$
for charged pions). Fitting the measured behaviour to two
exponentials we would find:
\begin{eqnarray}
\langle0|\pi^0(t)\pi^0(t)\,\sigma(0)|0\rangle &=&
A_{00}\exp(-E_0t)+B_{00}\exp(-E_1t)+\cdots\\
\langle0|\pi^+(t)\pi^-(t)\,\sigma(0)|0\rangle &=&
A_{+-}\exp(-E_0t)+B_{+-}\exp(-E_1t)+\cdots,
\end{eqnarray}
where, at tree-level in chiral perturbation theory, $E_0=2m_\pi$
and $E_1=2\sqrt{m_\pi^2+\vec{p}^{\ 2}_{\pi_\pm}}$. The ellipses
represent terms with higher energies and we assume here these can
be neglected. By fitting the correlation functions above, the
constants $A_{00},A_{+-},B_{00},B_{+-}$ can, in principle at
least, be determined numerically and we would then know which
combinations of the two-pion operators have no overlap with states
with energies $E_0$ and $E_1$ respectively (we call these states
$|s_0\rangle$ and $|s_1\rangle$ and denote by $\Pi^2_0$ and
$\Pi^2_1$ the operators with no overlap with $|s_1\rangle$ and
$|s_0\rangle$ respectively). Note that in order to extract
$B_{00}$ and $B_{+-}$ it is necessary to include the non-leading
exponential in the fit, which eliminates a major potential
advantage of using twisted boundary conditions for $K\to\pi\pi$
decays. Combining the results from these fits, together with those
of four-pion correlation functions of the form $\langle
0|\pi(t)\pi(t)\,\pi(0)\pi(0)|0\rangle$, we can determine the
matrix elements $\langle 0|\Pi^2_{0}|s_0\rangle$, $\langle
0|\Pi^2_{1}|s_1\rangle$, $\langle 0|\sigma|s_0\rangle$ and
$\langle 0|\sigma|s_1\rangle$. Unfortunately, even if we are able
to carry out the procedure described above with reasonable
accuracy, it is still not clear how to relate the
\emph{finite-volume} eigenstates $|s_0\rangle$ and $|s_1\rangle$
(which have different energies) to the \emph{infinite-volume}
eigenstates $|(\pi \pi)_{I=0}\rangle$ and $|(\pi
\pi)_{I=2}\rangle$ since the known procedures for doing this
\cite{ml,ll,lmst} rely on isospin symmetry. In some respects this
problem resembles the one of extending the discussion of
refs.~\cite{ml,ll,lmst} above the $K\bar{K}$ threshold. We
conclude that new ideas would be necessary before $K\to\pi\pi$
matrix elements could be determined with twisted boundary
conditions.

In order to overcome the above difficulties one should introduce
twisted boundary conditions which preserve isospin symmetry and
this is not possible in general.  Christ and Kim~\cite{kc}
however, have pointed out that one can make some progress for
$(\pi\pi)_{I=2}$ states if one restricts the twist angle to $\pi$
(they call this case $H$-parity). The $K^+\to\pi^+\pi^0$ matrix
element can be related by the Wigner-Eckart theorem to a matrix
element of a $\Delta I_z=3/2$ operator into a $\pi^+\pi^+$ final
state. By choosing $\vec{\theta}=(\pi,0,0)$ for the down quark and
$\vec{0}$ for the up quark and performing the Fourier transforms
over the positions of the two pions with weights 1 and
$\exp\{i(-2\pi/L)x\}$ respectively, we obtain a $\pi^+\pi^+$ final
state with the two pions having momenta $\pi/L$ and $-\pi/L$ in
the $x$ direction (hence remaining in the centre-of-mass frame).
With this procedure we are restricted to $\theta_x=\pi$ but the
need for extracting terms corresponding to non-leading
exponentials is avoided. Note that this procedure is possible,
because the required matrix element can be related to one in which
the final state only contains $\pi^+$ mesons. For $I=0$ final
states this is not possible, although in ref.~\cite{kc} it is also
shown that by introducing discrete ($G$-parity) boundary
conditions one can treat $I=0$ two-pion states with the two pions
having momenta $\pm\pi/L$ (but additional ideas will have to be
introduced to incorporate kaon states at rest into the formalism).

\section{Partially Twisted Boundary
Conditions}\label{sec:partiallytwisted}

Until now we have assumed that the twisted boundary conditions are
applied consistently to both the valence and sea quarks. In
lattice simulations this implies that a new set of gauge
configurations must be generated for each choice of the twist. In
addition, if different twists are imposed on the $u$- and
$d$-quark fields then one must use formulations of lattice
fermions for which the light quark determinant is positive
definite for each flavour. It would clearly be very welcome if one
could avoid new simulations for every value of $\vec{\Theta}$ and
in this section we analyse the consequences of introducing
different boundary conditions for sea and valence quarks. In
particular we consider the case in which the valence quarks
satisfy twisted boundary conditions and the sea quarks satisfy
periodic boundary conditions. In this case the QCD Lagrangian can
be conveniently written as: \beq {\cal L}=\bar q_v(x) \l ( \slash
\hspace{-7pt} \tilde{D}_v + M_v \r ) q_v(x)
    +\bar q_g(x) \l (\slash \hspace{-7pt} \tilde{D}_g + M_g \r ) q_g(x)
    +\bar q_s(x) \l (\slash \hspace{-7pt} \tilde{D}_s + M_s \r ) q_s(x)
    \label{pqQCD}
\eeq where the subscripts ${v,g,s}$ stand for valence, ghost and
sea and $q_g$ are commuting spinors. Moreover in order to have a
cancellation of valence loops we require that $D_g=D_v$ and
$M_v=M_g$. Eq.~(\ref{pqQCD}) can be rewritten in the form \beq
{\cal L}=\bar Q(x) \l ( \slash \hspace{-7pt} D + M \r ) Q(x)\,,
\quad\textrm{where}\quad Q(x)=(q_v(x),q_g(x),q_s(x))\,, \eeq and
now both $B_\mu$ and $M$ take values in the graded algebra of
$U(N_v+N_s|N_v)_V$~\footnote{Note that globally the structure of
the graded symmetry group is more involved \cite{shsh} but this is
not relevant for our discussion.}.

The derivation of the Feynman rules for both QCD and $\chi PT$ is
standard and we do not present it here. Having different twists
for valence and sea quarks breaks the valence-sea symmetry. This
is clearly a finite-volume effect, but the relevant question is
whether the corrections induced by this asymmetry decrease like
powers of the volume or exponentially\footnote{Again, low energy
constants are not affected by the twist as can be seen combining
the discussion made above for the full case with the one in
ref.~\cite{shsh2}.}. We find that the situation is analogous to
the violation of unitarity in partially quenched QCD~\cite{pq} and
that for many physical quantities (including those with at most a
single hadron in the initial and final states) the use of
partially twisted boundary conditions induces errors which are
exponentially small. This is because sea quarks appear in loops
and the sums over the loop-momenta can be approximated by
integrals with exponential precision.

We start by considering processes with at most one hadron in the
external states. As long as the shift does not induce cuts in the
correlation function, the correction is still exponentially
suppressed in the volume. For example, if we consider an
unquenched simulation with three flavours, in the asymptotic
limit, we find (in the notation defined in
eq.\,(\ref{eq:xdef})\,): \beq \frac{\Delta
f_{K^{\pm}}}{f_{K^{\pm}}}\to \l\{ \barr{c}
 - \frac94 \frac{m_\pi^2}{f_\pi^2}\frac{e^{-m_\pi L}}{(2\pi m_\pi
 L)^{3/2}}\hspace{2in}
 \textrm{(a)}\\ \\
 - \frac{m_\pi^2}{f_\pi^2}\frac{e^{-m_\pi L}}{(2\pi m_\pi L)^{3/2}}
    \l (\frac12\sum_{i=1}^3\cos{\theta_i}+\frac34\r)\hspace{0.85in}
 \textrm{(b)}\\ \\
 - \frac{m_\pi^2}{f_\pi^2}\frac{e^{-m_\pi L}}{(2\pi m_\pi L)^{3/2}}
    \l (\sum_{i=1}^3\cos{\theta_i}-\frac34\r)\hspace{1in}
 \textrm{(c)}
\earr \r. \nn \eeq for the three cases in which the $d$ and $s$
quarks satisfy periodic boundary conditions but the up quark is
(a) untwisted, (b) fully twisted (both valence and sea $u$-quarks
satisfy twisted boundary condition) and (c) partially twisted
(only the valence $u$ quark is twisted). This shows that, in
general, finite volume corrections could be different for the
three cases but they are always exponentially small.

In \cite{gp} Golterman and Pallante demonstrated that (partial)
quenching can induce ``unphysical" mixing among weak operators
because of their different transformation properties under the
flavour group and its graded extension. These effects are
proportional to the difference $M_s-M_v$ and have to disappear in
full QCD. One could also ask whether imposing different boundary
conditions for sea and valence quarks could lead to similar
effects. Such effects are proportional to $\theta_v-\theta_s$ and
again appear as exponentially small finite-volume corrections.

Not surprisingly the case of amplitudes with multiparticle
external states is much more complicated. We have seen in
sec.\,\ref{sec:fveffects} that it is not possible to isolate
$\pi\pi$ states with a given isospin using periodic boundary
conditions. We therefore restrict our consideration here to the
H-parity and G-parity cases for $I=2$ and $I=0$ two-pion states
respectively. Since now the twist angle is fixed to be $\pi$, the
practical advantage of using partial twisting to avoid generating
new gluon configurations for every value of the angle is much less
compelling, but it is interesting nevertheless to consider the
theoretical issues. The effects of different boundary conditions
for sea and valence quarks are analogous to those discussed in
\cite{pq} for the partially quenched theory. For $I=2$ the sea
quarks do not enter the FSI directly and the difference between
imposing H-parity boundary conditions fully or partially is
exponentially small. (The power corrections in the volume with
H-parity boundary conditions are different of course from those
with periodic ones, but they are calculable~\cite{dt,rg}.) For
$I=0$ and partial $G$-parity boundary conditions on the other
hand, the mesons in intermediate states in correlation functions
necessarily include both sea and valence quarks, whereas the
external states are made of valence quarks only. The lack of
degeneracy between external and internal states implies a
breakdown of unitarity and Watson's theorem and we are then unable
to extract the physical matrix elements.

In conclusion we have found that for a large class of processes
(those without final state interactions) it is possible to neglect
the twist of the determinant avoiding the need to generate new
gauge configurations for each twist. This is not true however, for
all processes. In particular, for $K\to\pi\pi$ matrix elements,
with the two pions in an $I=0$ state, if G-parity boundary
conditions are used they must be implemented for both the valence
and sea quarks.

\section{Conclusions}\label{sec:concs}
In this letter we have used $\chi$PT to study the finite-volume
corrections with twisted boundary conditions. For quantities
without final-state interactions, such as meson masses, decay
constants or semileptonic and other form-factors, we confirm that
these corrections remain exponentially small in the volume. This
remains true with partially twisted boundary conditions for which
only the valence quarks are twisted, thus eliminating the need to
generate a new set of gluon configurations for each choice of
twisting angle and makes the technique much more useful.

We have also demonstrated that twisted boundary conditions cannot
be applied in general to processes with final-state interactions,
such as $K\to\pi\pi$ decays. This is disappointing since twisted
boundary conditions would have been particularly useful for
lattice studies of these decays, extending very significantly the
kinematic range accessible in a simulation. In spite of this
particular disappointment, we look forward to the implementation
of twisted boundary conditions to the wide range of processes for
which they are applicable and phenomenologically useful.

\subsection*{Acknowledgments}
We warmly thank Guido Martinelli, Mauro Papinutto and Steve Sharpe
for helpful discussions and Maarten Golterman and Santi Peris for
their hospitality during the Benasque workshop on \textit{Matching
Light Quarks to Hadrons} where this work was initiated. CTS
acknowledges support from PPARC grant PPA/G/O/2002/00468 and GV
acknowledges support from the European Commission under the RTN
contract MRTN-CT-2004-503369 (Quest for Unification).

\appendix
\section{Finite-Volume Corrections in Chiral Perturbation Theory}
\label{app:tadpoles}

In this appendix we derive the finite-volume corrections with
twisted boundary conditions at one-loop order in $\chi$PT. The
generic expression for tadpole diagrams in finite volume is given
by the left-hand side of
\begin{equation} \frac{1}{L^3}\sum_{\vec
q} \frac{1}{\l[\l(\vec q + \frac{\vec \theta}{L}\r)^2+M^2\r]^s}=
\frac{\sqrt{4 \pi}\ \Gamma(s+\frac12)}{\Gamma(s)} \int \frac{d^4
q}{(2\pi)^4} \frac{1}{(q^2+M^2)^{s+\frac12}}+ \xi^\theta_s(L,M)
\,. \label{eq:defxi} \end{equation} The first term on the
right-hand side of eq.\,(\ref{eq:defxi}) is the corresponding
infinite-volume integral and $\xi^\theta_s(L,M)$ contains the
finite-volume corrections. We now generalise the procedure of
ref.\,\cite{dg} to twisted boundary conditions and demonstrate
that these corrections are exponentially small in the volume.
\begin{eqnarray}
\xi^\theta_s(L,M)&=& \frac{1}{L^3}\sum_{\vec q}
\frac{1}{\l[\l(\vec q+ \frac{\vec \theta}{L}\r)^2+M^2\r]^s}-
\frac{\sqrt{4 \pi}\ \Gamma(s+\frac12)}{\Gamma(s)} \int \frac{d^4
q}{(2\pi)^4} \frac{1}{(q^2+M^2)^{s+\frac12}}\cr &=&
\frac{1}{\Gamma(s)} \int_0^\infty d\tau\, \tau^{s-1} e^{-\tau M^2}
\frac{1}{L^3}\sum_{\vec q} e^{-\tau (\vec q+\vec \theta/L)^2} -
\frac{1}{\Gamma(s)} \int_0^\infty d\tau \tau^{s-1} e^{-\tau
M^2}\int {d^3q\over (2\pi)^3} e^{-\tau \vec q^2}\cr
&=&\frac{1}{\Gamma(s)}\int_0^{\infty}d\tau\, \tau^{s-1}e^{-\tau
M^2}
\l[\frac{1}{L^3}\prod_{i=1}^3\vartheta\l(\frac{4\pi^2\tau}{L^2},
\frac{\theta_i}{2\pi}\r)-\frac{1}{8(\pi \tau)^{3/2}} \r]\label{e1} \\
&=&\frac{L^{2s-3}}{(2\pi)^{2s}\Gamma(s)}\int_0^{\infty} d\tau\,
\tau^{s-1} e^{-\tau \l(\frac{ML}{2\pi}\r)^2}
\left[\prod_{i=1}^3\vartheta\l(\tau,\frac{\theta_i}{2\pi}\r)-\l(\frac{\pi}{\tau}\r)^{3/2}\right]
\label{e1b} , \eea where we have defined the elliptic
$\vartheta$--function $\vartheta(\tau,\alpha)$ by: \beq
\vartheta(\tau,\alpha)\equiv\sum_{n=-\infty}^{\infty}e^{-\tau(n+\alpha)^2}\,.
\eeq $\vartheta(\tau,\alpha)$ satisfies the Poisson summation
formula:
\begin{equation}
\vartheta(\tau,\alpha)=\sqrt{\frac{\pi}{\tau}}e^{-\tau\,\alpha^2}
\vartheta\l ( \frac{\pi^2}{\tau},-i\frac{\alpha \tau}{\pi}\r) \eeq
so that \bea \xi^\theta_s(L,M)&=&
\frac{1}{(4\pi)^{3/2}\Gamma(s)}\int_0^{\infty}d\tau\
\tau^{s-5/2}e^{-\tau M^2} \l[
\prod_{i=1}^3\vartheta\l(\frac{L^2}{4\tau},-i\frac{2\theta_i
\tau}{L^2}\r)e^{-\tau\theta_i^2/L^2}-1  \r]. \eea The leading
finite-volume corrections are now readily obtained. Using
\begin{equation} \vartheta\l(\frac{L^2}{4\tau},-i\frac{2\theta_i
\tau}{L^2}\r)e^{-\tau\theta_i^2/L^2}
=\sum_{m=-\infty}^{\infty}e^{-\frac{L^2}{4\tau}m^2+i\theta_i m}
=\sum_{m=-\infty}^{\infty}e^{-\frac{L^2}{4\tau}m^2}\cos(\theta_i
m) \label{eq:thetaexp}\end{equation} we see that for large $L$
\begin{equation}
\vartheta\l(\frac{L^2}{4\tau},-i\frac{2\theta_i
\tau}{L^2}\r)e^{-\tau\theta_i^2/L^2}\rightarrow
1+2e^{-\frac{L^2}{4\tau}}\cos(\theta_i)\,. \end{equation} If
$\cos(\theta_i)=0$, then the leading finite-volume corrections are
given by the $m=\pm 2$ terms in eq.\,(\ref{eq:thetaexp}) and hence
decrease with a larger exponent. For the generic case in which
$\cos(\theta_i)\neq 0$ for $i=1,2,3$ the behaviour of the
finite-volume corrections as $L\to\infty$ is given by
\begin{eqnarray}
\xi_s^\theta(L,M)&\to&
\frac{\sqrt{\pi}}{\Gamma(s)(2\pi)^{3/2}}\frac{e^{-ML}}{(ML)^{2-s}}(2M^2)^{3/2-s}
(\cos\theta_1+\cos\theta_2+\cos\theta_3)\label{eq:expsmall}\\
&=&\xi_s^0(L\to\infty,M)\times\frac{\cos\theta_1+\cos\theta_2+\cos\theta_3}{3}\,,
\eea where $\xi_s^0$ are the finite-volume corrections with
$(\theta_1,\theta_2,\theta_3)=\vec 0$. Eq.\,(\ref{eq:expsmall})
demonstrates that finite-volume corrections are exponentially
small with twisted boundary conditions.

The second diagram which appears at one loop level contains two
propagators and in finite volume gives a contribution proportional
to: \beq \Bigl. \frac{1}{L^3}\sum_{{\vec k}}\int\frac{d k_4}{2\pi}
\frac{{\cal N} }{[(k+A_1)^2+m_1^2][(q-k+A_2)^2+m_2^2]}\,,
\label{eq:2part} \eeq where the numerator $\cal N$ is a function
of momenta and masses and  $q$ is the injected momentum. For
illustration we consider the simplest case for which ${\cal N}=1$
(terms in ${\cal N}$ containing the loop momentum can be reduced
to the tadpole integrals of the form in eq.\,(\ref{eq:defxi})\,)
and $\vec q=0$ (the finite-volume effects in integrals with
non-zero $\vec{q}$ can readily be obtained from the expressions
below by the substitution $\vec{A_2}\to\vec{A_2}+\vec{q}$\,).

After introducing the Feynman parameter $x$ and performing the
$k_4$ integration eq.~(\ref{eq:2part}) reduces to:
\beq\label{eq:2part2} \frac14\int_0^1 dx\,\frac{1}{L^3}\sum_{\vec
k}\l[\l(\vec k+\vec A(x)\r)^2+M^2(x)\r]^{-3/2}\,, \eeq where \bea
\vec A(x)&=&\frac{\vec\theta(x)}{L}=x \vec A_1-(1-x) \vec A_2\,, \nn \\
M^2(x)&=&(1-x)m_2^2+x m_1^2+x(1-x)(q^2+(\vec{A}_1+\vec{A}_2)^2)\nn \\
&=&(1-x)m_2^2+x m_1^2-x(1-x)\l(E^2-(\vec A_1+\vec A_2)^2\r)\,,\nn
\eea where in the last line we have made the replacement $q_4\to
iE$ and $E$ is the physical (Minkowski) injected energy.
Eq.\,(\ref{eq:2part2}) has the same form as the left-hand side of
eq.~(\ref{eq:defxi}), so we can proceed just as for the tadpole
integral to obtain the expression for the corresponding
$\xi^\theta$-function: \beq
\xi^\theta_{3/2}(L,m_1,m_2,q)=\frac{2}{ (2 \pi)^{3}\sqrt{\pi}
}\int_0^1 dx\,\int_0^{\infty} d\tau\, \tau^{1/2} e^{-\tau
\l(\frac{M(x) L}{2\pi}\r)^2}
\left[\prod_{i=1}^3\vartheta\l(\tau,\frac{\theta_i(x)}
{2\pi}\r)-\l(\frac{\pi}{\tau}\r)^{3/2}\right]\,, \eeq which is
exponentially small in the volume as long as $M^2(x)>0$, i.e. as
long as no branch cuts appear.

In partially quenched chiral perturbation theory there are also
contributions with \textit{double poles} in which one or both
propagators in eq.\,(\ref{eq:2part}) are squared. These can be
written in terms of derivatives of (\ref{eq:2part}) w.r.t. the
masses and the finite-volume corrections therefore remain
exponentially small.

\section{Masses and Decay Constants} \label{app:fvfull}

In this appendix we present the full finite-volume corrections for
meson masses and decay constants at NLO in $\chi$PT with twisted
boundary conditions (using the notation of
eq.\,(\ref{eq:xdef})\,): \bea \frac{\Delta
m_{\pi^{\pm}}^2}{m_{\pi^{\pm}}^2}&=&
    \frac{1}{2 f^2}\xi_{1/2}(L,m_{\pi^0})-\frac{1}{6 f^2}\xi_{1/2}(L,m_{\eta})\,, \nn \\
\frac{\Delta m_{\pi^{0}}^2}{m_{\pi^{0}}^2}&=&
    \frac{1}{f^2}\xi^{\theta}_{1/2}(L,m_{\pi^\pm})-\frac{1}{2 f^2}\xi_{1/2}(L,m_{\pi^0})-\frac{1}{6 f^2}\xi_{1/2}(L,m_{\eta})\,, \nn \\
\frac{\Delta m_{K^{\pm}}^2}{m_{K^{\pm}}^2}&=&
    \frac{1}{3 f^2}\xi_{1/2}(L,m_{\eta})\,, \nn \\
\frac{\Delta m_{K^{0}}^2}{m_{K^{0}}^2}&=&
    \frac{1}{3 f^2}\xi_{1/2}(L,m_{\eta})\,, \nn \\
\frac{\Delta f_{\pi^{\pm}}}{f_{\pi^{\pm}}}&=&
    -\frac{1}{2 f^2}\xi^{\theta}_{1/2}(L,m_{\pi^\pm})-\frac{1}{2 f^2}\xi_{1/2}(L,m_{\pi^0})
    -\frac{1}{4 f^2}\xi^{\theta}_{1/2}(L,m_{K^\pm})-\frac{1}{4 f^2}\xi^{\theta}_{1/2}(L,m_{K^0})\,, \nn \\
\frac{\Delta f_{\pi^{0}}}{f_{\pi^{0}}}&=&
    -\frac{1}{f^2}\xi^{\theta}_{1/2}(L,m_{\pi^\pm})
    -\frac{1}{4 f^2}\xi^{\theta}_{1/2}(L,m_{K^\pm})-\frac{1}{4 f^2}\xi^{\theta}_{1/2}(L,m_{K^0})\,, \nn \\
\frac{\Delta f_{K^{\pm}}}{f_{K^{\pm}}}&=&
    -\frac{1}{4 f^2}\xi^{\theta}_{1/2}(L,m_{\pi^\pm})-\frac{1}{8 f^2}\xi_{1/2}(L,m_{\pi^0})-\frac{3}{8 f^2}\xi_{1/2}(L,m_{\eta})
    -\frac{1}{2 f^2}\xi^{\theta}_{1/2}(L,m_{K^\pm})-\frac{1}{4 f^2}\xi^{\theta}_{1/2}(L,m_{K^0})\,, \nn \\
\frac{\Delta f_{K^{0}}}{f_{K^{0}}}&=&
    -\frac{1}{4 f^2}\xi^{\theta}_{1/2}(L,m_{\pi^\pm})-\frac{1}{8 f^2}\xi_{1/2}(L,m_{\pi^0})-\frac{3}{8 f^2}\xi_{1/2}(L,m_{\eta})
    -\frac{1}{4 f^2}\xi^{\theta}_{1/2}(L,m_{K^\pm})-\frac{1}{2 f^2}\xi^{\theta}_{1/2}(L,m_{K^0})\,, \nn
\eea where the $\xi_s^\theta(L,m_f)$-functions are defined in
appendix~\ref{app:tadpoles} and the twist angle
$\theta=\theta_{f}$ is the one associated with the meson of
flavour $f$ (e.g. $\theta_{\pi^+}=\theta_u-\theta_d$).

\end{document}